# Are Classical Clone Detectors Good Enough For the AI Era?

Ajmain I. Alam, Palash R. Roy, Farouq Al-omari[†], Chanchal K. Roy, Banani Roy, Kevin A. Schneider
Department of Computer Science, University of Saskatchewan, Saskatoon, Canada
[†]Department of Engineering, Thompson Rivers University, Kamloops, Canada
E-mail: {ajmain.alam, palash.roy, chanchal.roy, banani.roy, kevin.schneider}@usask.ca, [†]falomari@tru.ca

*Abstract*—The increasing adoption of AI-generated code has reshaped modern software development, introducing syntactic and semantic variations in cloned code. Unlike traditional human-written clones, AI-generated clones exhibit systematic syntactic patterns and semantic differences learned from large-scale training data. This shift presents new challenges for classical code clone detection (CCD) tools, which have historically been validated primarily on human-authored codebases and optimized to detect syntactic (Type 1–3) and limited semantic clones. Given that AI-generated code can produce both syntactic and complex semantic clones, it is essential to evaluate the effectiveness of classical CCD tools within this new paradigm. In this paper, we systematically evaluate nine widely used CCD tools using GPTCloneBench, a benchmark containing GPT-3-generated clones. To contextualize and validate our results, we further test these detectors on established human-authored benchmarks, BigCloneBench and SemanticCloneBench, to measure differences in performance between traditional and AI-generated clones. Our analysis demonstrates that classical CCD tools, particularly those enhanced by effective normalization techniques, retain considerable effectiveness against AI-generated clones, while some exhibit notable performance variation compared to traditional benchmarks. This paper contributes by (1) evaluating classical CCD tools against AI-generated clones, providing critical insights into their current strengths and limitations; (2) highlighting the role of normalization techniques in improving detection accuracy; and (3) delivering detailed scalability and execution-time analyses to support practical CCD tool selection. The research underscores the continued relevance of classical CCD tools and suggests adopting a hybrid approach that combines both classical and AI-based methods to improve clone detection in the modern era.

*Index Terms*—Software Clone, Code Clone Detection, Tool Evaluation, AI

## I. Introduction

Code clones, which are identical or similar code fragments, play a significant role in software maintenance and evolution [49], [6]. They often arise when developers reuse code through copying, pasting, and modifying it within or across projects, which can lead to code decay and maintenance challenges, making code clone detection (CCD) critical for ensuring software quality. Numerous CCD tools and techniques have been developed over the years, and in particular, classical CCD tools have benefited significantly from extensive research and practical applications [48], [62]. These tools effectively detect clones by leveraging structural similarity analysis and have been shown to scale efficiently to large datasets on standard hardware without modification [62]. Additionally, they are fast, adaptable, and widely adopted, making them practical for real-world software development. However, their performance remains limited when it comes to detecting Type-4 or semantic clones, highlighting a critical gap in their capabilities.

AI-based software clone detection tools have been introduced to address the limitations of classical clone detectors in identifying semantic clones. A recent survey [37] reported that over 20 AI-based clone detectors have been proposed since 2016, indicating a growing preference in the code clone community for AI-driven techniques, especially for detecting semantic clones. AI-based detectors have shown superior capabilities in capturing deeper semantic relationships beyond syntactic and structural similarities compared to classical methods. However, these advantages come at a cost. AI-based detectors often lack interpretability, making it difficult to trust or validate their results. They also require large and diverse training datasets and are computationally intensive, limiting their practicality in real-time or resource-constrained environments [73], [72]. Moreover, it remains uncertain whether their enhanced performance is due to their architecture or potential biases in the training data, highlighting the need for further investigation.

The widespread adoption of AI-assisted software development coding tools has fundamentally changed software development. A recent survey shows that 92% of North American developers now use AI-assisted tools in their workflows [21], with major technology companies such as Google integrating AI-powered code generation into their development pipelines [30]. As a result, AI-generated code is no longer an experimental novelty, it has become a standard part of modern software systems. Despite this shift, classical CCD tools are the industry standard for code similarity analysis and are widely adopted due to their efficiency, scalability, and reliability [64], [14]. However, since the AI boom, no systematic evaluation has been conducted to assess whether these classical tools can handle AI-generated code effectively. Given the increasing presence of AI-generated code in real-world software projects, it is critical to re-evaluate classical clone detectors in this new context.

To assess the effectiveness of classical clone detectors on AI-generated code, we employ a structured evaluation framework. CCD tools are traditionally assessed based on precision, recall, execution time, and scalability [62]. Precision is measured by manually verifying a sample of reported clones, while execution time and scalability are evaluated by applying tools to software systems of varying sizes. However, recall evaluation presents a greater challenge as it requires a benchmark of known reference clones [48], [56]. Constructing these benchmarks involves accurately injecting test clones and developing robust recall evaluation mechanisms [2], [46], [55]. While some studies omit recall evaluation and rely solely on precision through manual validation [62], [20], [28], [31], [32], [39], the rise of AI-generated code necessitates a more rigorous approach.

Benchmarks have played a crucial role in advancing CCD research. The foundational Bellon et al. benchmark [8] enabled early tool comparisons, while BigCloneBench [54] introduced an automated framework (BigCloneEval) for recall measurement [57]. More recently, SemanticCloneBench [2] was developed to evaluate tools on semantic clones. The latest development, GPTCloneBench, leverages GPT-3-generated clones [3], providing an opportunity to assess the effectiveness of classical clone detectors in handling AI-generated clones.

To provide a comprehensive assessment, we evaluate a diverse range of CCD tools, not only to test their performance but also to serve as a performance baseline for understanding the limitations of syntax-based techniques. We aim to determine whether classical CCD tools can still offer meaningful benefits in handling AI-generated clones. However, classical CCD tools primarily depend on benchmarks like BigCloneBench [59], which do not accurately reflect the characteristics of AI-generated clones. To address this gap, we evaluate them on GPTCloneBench [3], alongside SemanticCloneBench [2], which contains human-written semantic clones. This comparison allows us to analyze performance variations between AI-generated and human-authored clones. Through this approach, our research not only fills a critical gap in software maintenance and development but also lays the groundwork for AI-aware CCD methodologies, providing deeper insight into the capabilities and limitations of classical CCD tools for AI-generated code. In this paper, we address the following research questions to address the gap:

**(RQ1)** How do classical clone detection tools perform on AI-generated clones?

**(RQ2)** Do tools perform differently in AI versus non-AI clone benchmarks?

**(RQ3)** What is the execution time and scalability for each tool?

**Replication Package is here** https://github.com/srlabUsask/classical-clone-detector-in-ai-era. The paper is organized as follows: Section II outlines the motivation. Section III defines key terms, and Section IV reviews related research. The experimental setup is in Section V, with results in Section VI and a discussion in Section VII. Threats to validity are in Section VIII, and Section IX concludes the paper.

## II. MOTIVATION

To discuss the motivation, we have highlighted the differences between human-generated and AI-generated code through preliminary examples and insights from expert opinions, emphasizing the need to systematically evaluate existing classical CCD tools in the context of AI-generated code.

### A. Human vs AI code

Software development field is rapidly evolving [50]. With the advancement of Large Language Models (LLMs), compare to human-written code, AI-generated code can exhibit unique patterns, idioms, structures, stylistic elements, and abstractions [27], [40], [26], [44], [11], [43]. Human-authored code can vary depending on individual preferences, and the specific requirements of a task. However, AI can combine knowledge from diverse sources that lead to novel or unorthodox approaches, which might not be common in traditional human design patterns. These contrasts underscore the need for re-examining how code similarity is evaluated in modern software ecosystems. Although existing CCD tools specially the classical ones, have proven effective with previous benchmarks, their performance and reliability on AI-generated code remain underexplored. As we are discussed earlier about AI-based agents and codes are becoming more common, evaluating and potentially refining these CCD tools to account for AI-specific properties is increasingly crucial. Recent studies further highlighted the limitations of traditional clone detectors in this AI-driven context. For instance, over 99% of AI-generated code snippets from models like ChatGPT managed to evade detection by MOSS, a widely-used CCD tool, raising concerns about the robustness of existing tools [63]. This shows that AI-generated code poses a new challenge for CCD tools.

### B. Example of Human vs AI Code

To concretely illustrate the challenges posed by AI-generated clones, we performed a preliminary comparative analysis between human-authored and AI-generated solutions for checking palindromes in Python (Figure 1). Human-generated solutions collected from Stack Overflow. **Human Answer 1** (Figure 1a) employed a concise implementation using a for loop combined with Python's built-in functions zip and reversed to directly compare corresponding characters from opposite ends of the string. **Human Answer 2** (Figure 1b) utilized a conventional for loop iterating only up to half of the string length, explicitly comparing mirrored characters.

By contrast, ChatGPT-generated answers demonstrated AI's potential for both replication and innovation. Notably, **ChatGPT Answer 1** (Figure 1c) closely replicated the syntactic structure of Human Answer 1 but introduced an enhancement addressing a previously unhandled corner case: converting the entire input string to lowercase when all characters were alphabetic. Such subtle yet meaningful modifications exemplify AI's ability to generate refined syntactic clones enriched with additional logic. **ChatGPT Answer 2** (Figure 1d) diverged significantly from both human examples by employing Python's deque structure, showcasing AI's capacity to produce semantic clones through alternative data structures and algorithmic choices.

This observation reveals that it is important to evaluate existing classical CCD tools systematically against AI-generated code. Such evaluations will clarify whether these tools can reliably identify clones produced by contemporary AI systems.

*C. Expert Opinion*

To validate the relevance of our research, we conducted semi-structured interviews with five software engineering experts, each with 8+ years of programming experience, 2+ years of professional practice, and direct experience implementing CCD tools in development environments. These interviews revealed critical insights about the changing nature of code clones in AI-assisted development workflows. All experts independently highlighted distinctive characteristics of AI-generated code compared to human-authored code: greater structural consistency, more comprehensive documentation, and stricter preference for formal coding guidelines. Three experts specifically noted that AI-generated code introduces a "detection blind spot" for traditional CCD tools, as these tools were calibrated against human coding patterns and variability. One senior developer stated that we're seeing false negatives in our clone detection pipeline that weren't there before our team started using AI assistants. Importantly, experts challenged the assumption that AI generates only semantic clones, noting that modern AI systems produce a spectrum of clone types depending on prompt framing, problem context, and implementation constraints. This observation directly supports our hypothesis that comprehensive reevaluation of classical CCD tools is necessary. Four of five experts expressed specific concerns about the lack of validation for existing CCD tools against AI-generated code variants, with all emphasizing the urgent need for rigorous assessment. As one software team lead remarked that we're flying blind when it comes to understanding how reliable our existing clone detection infrastructure is in this new development paradigm. These expert perspectives align with emerging research suggesting that CCD methodologies must evolve to maintain effectiveness within increasingly AI-influenced software ecosystems. The full interview transcripts and other details can be made available upon request.

## III. DEFINITIONS

**Type-1 (T1):** Clones represent code fragments that are syntactically identical, differing only in white space, layout, and comments [48], [8], [47], [57].

**Type-2 (T2):** Clones correspond to code fragments that are syntactically identical, with variations in identifier names and literal values, in addition to the Type-1 clone differences [48], [8].

**Type-3 (T3):** Clones encompass syntactically similar code fragments that diverge at the statement level. These fragments involve the addition, modification, and/or removal of statements in relation to each other, in addition to the Type-1 and Type-2 clone differences [48], [8], [3].

**Type-4 (T4):** Code fragments that are syntactically dissimilar but serve the same functional purpose [48], [2], [3]. With the absence of a universally agreed-upon threshold for the minimum syntactical similarity required to classify a clone as Type-3, the differentiation between Type-3 and Type-4 clone pairs that embody the same functionality, as exemplified in BigCloneBench [53], presents a complex conundrum.

In our study, we have followed the categorization scheme of BigCloneBench [57], [53] for code clones based on their similarity percentage. These categories encompass Moderately Type-3 clones (MT3) falling in the 50-70% range, and Weakly Type-3/Type-4 clones (WType-3/Type-4) exhibiting a similarity range of 0-50% [53], [56], [3]. On the other hand, GPTClonebench authors also followed the same categorization techniques but with an additional 5% for each category named as a grey area for MT3, making the similarity range 50%-75%.

## IV. RELATED WORK

Benchmarks are established through various methods, including manual clone extraction from systems [53], [34], manual validation of tool-generated results for specific subject systems [13], [9], [18], or intentional introduction of known clones into a system [61], [46], [55], [56]. Bellon's Benchmark [8] evaluated the relative recall of six contemporary tools across C and Java subject systems, manually validating 2% of the 325,935 potential clones identified. Murakami et al. [41] expanded the benchmark, addressing gaps in Type-3 reference clones and enhancing clone-matching metrics [55], [56], [57]. However, Bellon's Benchmark has biases, with a growing gap as CCD techniques advance, especially in Type-3 detection. Incorporating clones from newer tools is challenging due to limited documentation on Bellon's validation process, as noted by Baker [7], [8], [55], [56].

The Mutation and Injection Framework [46], [53], introduced by Svajlenko and Roy, stands as a synthetic clone benchmark and is designed to generate customizable benchmarks consisting of artificial copy-and-paste clones using a mutation-analysis process. It has the ability to evaluate recall for individual edit types, offering a level of granularity beyond that of clone types. It offers the advantage of delivering unbiased, controlled, and finely detailed assessments of a tool's capabilities. Their research has demonstrated its effectiveness when applied with modern tools, offering profound insights into their clone recall performance [55], [56]. In Svajlenko and Roy's work [55], they conducted evaluations of eleven distinct tools, utilizing both Bellon's Benchmark and their in-house Mutation Framework, and identified inconsistencies in the recall measurements obtained from Bellon's Benchmark. As a result, they put forward the notion that Bellon's Benchmark may not be well-suited for assessing modern tools [56]. For this reason, Svajlenko and Roy introduced BigCloneBench[53], a contemporary real-world clone benchmark that comprises manually validated clones. The creation of BigCloneBench involved an extensive process of mining thousands of software systems to identify clones encompassing 43 different functionalities.

Machine learning (ML) has recently gained attention for code clone and similarity identification, with BigCloneBench

```python
def ispalindrome(s):
    for c1, c2 in zip(s, reversed(s)):
        if c1 != c2:
            return False
    return True
```
(a) Human Answer 1 [42], [16]

```python
def ispalindrome(s):
    for i in range(len(s)//2):
        if s[i] != s[-(i+1)]:
            return False
    return True
```
(b) Human Answer 2 [42], [16]

```python
def is_palindrome(s):
    s = ''.join(char.lower() for char in s if char.isalnum())
    return all(a == b for a, b in zip(s, reversed(s)))
```
(c) ChatGPT Answer 1

```python
from collections import deque
def is_palindrome(s):
    d = deque(s)
    while len(d) > 1:
        if d.popleft() != d.pop():
            return False
    return True
```
(d) ChatGPT Answer 2

Fig. 1: Palindrome solutions in Python from Human and ChatGPT. The first two images (1a, 1b) represent human-generated answers collected from Stack Overflow, while the last two images (1c, 1d) are generated by ChatGPT.

initially designed to evaluate modern CCD tools finding extensive use in training ML models for semantic CCD [59], [60]. Despite its primary purpose, BigCloneBench presents challenges when employed as a training dataset due to design issues that lead to imbalances and ambiguities in semantic clone definitions [33]. ML-based detectors trained on BigCloneBench may overlook semantic clones or yield inaccurate results [60], [33]. Additionally, BigCloneBench only includes Java clones, limiting its applicability to other languages. Addressing these concerns, Al-omari introduced SemanticCloneBench, which contains 4,000 Type-4 clones but is considered small for ML model training and testing. To address this limitation, Alam and Roy et al. introduced GPTCloneBench, an AI-generated benchmark featuring semantic and cross-language clone pairs for four programming languages [2], [3]. GPTCloneBench serves as a major tool to evaluate current clone detectors in the AI era, offering valuable insights into their efficiency in detecting AI-generated clones. Even though there are evaluation studies [56], [57], [1], [47], [52], none of them extensively focused on evaluating tools on AI-generated codes and comparing the performance with existing benchmarks.

## V. EXPERIMENTAL SETUP

Our approach focuses on making evaluation automated which is controlled, and free from bias.

### A. Injection Framework

The framework we introduced has two phases (Figure 2). The first phase is evolving the system phase, where we take a system and evolve it by injecting clones for the clone detector tools to search. This is similar to the injection approach provided in BigCloneEval and Mutation-injection framework [57], [61], [46]. The second phase is executing different CCD tools for the evolved system and analyzing their output to measure recall for the benchmark clone only. In the first phase, we evolve a system by taking real clones by mining the functions from a particular benchmark. Here, we give a system to the framework, and the framework finds two locations to inject a single clone. After that, we select one clone pair from the benchmark and inject two code fragments from that pair into the selected locations. During the evaluation phase, the tools are executed for each of the evolved systems. *For the selection of the subject systems, we used the same subject systems mentioned in GCB and SCB, which are JHotDraw, Django, PostgreSQL, and Mono.* Later, for RQ3, we modified the files based on the lines of code required for RQ3. The framework's extensibility is highlighted, as any tool can be integrated into the system by implementing a tool runner, a straightforward communication protocol enabling interaction between the framework and the tool. The performance of the tool is assessed for each evolved system, with a focus on unit recall. Unit recall is a measure of how well a tool performs. It's either 1 if the tool correctly finds the added clone or 0 if it doesn't. After every cycle, the framework reset the evolved system by removing the injected clone and starting a new cycle. For each clone pair in the benchmark, the same cycle is repeated by evolving the system with a new clone pair, running the CCD tool, and evaluating the tools' results. We have utilized this injection process on those tools that provided support to detect clones in a system. The way we designed our overall framework and experiment where our focus was never to extend a tool to provide more support. Our experimental framework was designed to use tools in their default state, focusing on inputting data, collecting outputs, and then analyzing these results without modifying the tools' capabilities. However, implementing this approach with tools like ASTNN and CodeBERT presented challenges, especially in terms of time efficiency; running these tools on a project with the specified configuration could exceed 100 days for a single result. Consequently, we did not apply our proposed injection process (outlined in Section V-A) to ASTNN and CodeBERT. Instead, we used stand-alone clone pairs from all benchmarks to detect clones with these tools, allowing us to evaluate their performance without the extensive time commitment that our injection process would require.

### B. Approach and Tool Configuration

**GPTCloneBench (GCB):** GPTCloneBench consists of 37,149 Moderate Type-3 and Type-4 clone pairs along with 20,770 cross-language clone pairs. In this experiment, we have only focused on the Moderate Type-3 and Type-4 clone pairs of GCB. We structured the execution of tools from stand-alone clones within GCB. In our evaluation process utilizing GCB, we adopt a coverage-based clone matching metric to confirm whether a reference clone within the benchmark is successfully detected by a candidate clone reported by a tool. This metric, briefly referred to as the coverage-match or c-match.

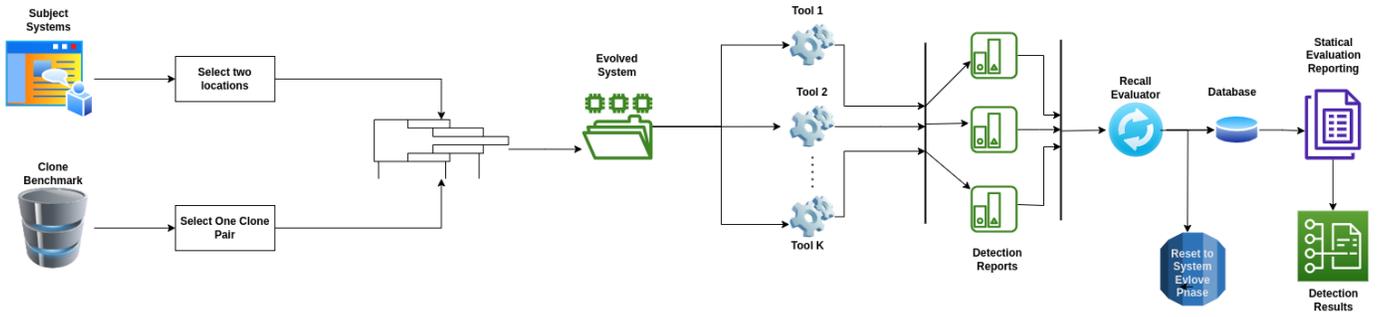

Fig. 2: Injection Framework

To explain, a code fragment, $f_1$, covers another fragment, $f_2$, if it intersects a fraction $t$ of $f_2$'s source lines within the same file (Equation 1). The c-match criterion defines a candidate clone $C$ as a match to a reference clone $R$ when $C$'s fragments sufficiently cover $t$ of $R$ (Equation 2). To accommodate variation in fragment ordering and potential off-by-one line differences among tools, the metric evaluates both orderings of the candidate clone's fragments. Consistent with prior works [8], [55], [56], we set the minimum coverage threshold $t$ to 70%, maintaining meaningful overlap. A tool's recall is then the ratio of benchmark reference clones matched by its reported candidate clones, as per the c-match metric.

$$covers(f_1, f_2, t) = \frac{min(f_1.e, f_2.e) - max(f_1.s, f_2.s) + 1}{(f_2.e - f_2.s) + 1} \geq t \quad (1)$$

$$c-match(C, R, t) = covers(C.f_1, R.f_1, t) \wedge covers(C.f_2, R.f_2, t) \quad (2)$$

**SemanticCloneBench (SCB):** In a similar approach to our GCB evaluation, we applied the same methodology to assess CCD tools using SemanticCloneBench. The SCB encompasses 4,000 clone pairs distributed across four programming languages: Python, Java, C, and C#. To mitigate potential scalability challenges, we have followed a similar approach to SCB by running the tools on a smaller project. This approach guaranteed comprehensive coverage of all clones within SCB.

**BigCloneBench (BCB):** We initiated our evaluation by executing the tools within the context of IJaDataset 2.0 source repository which crawled from SECold Project [23], meticulously calculating their recall performance regarding the clones contained in BigCloneBench.

**Choices of Subject Tools:** Many CCD tools and techniques are available in the literature. Evaluating all tools is a tedious task. Therefore, we selected open-sourced, popular and easy-to-replicate tools from tools' categories by following similar categorization of [25]. We followed a common procedure that a user or developer follows in finding a tool for a certain task. Even though our main focus was on non-AI-based classical detectors since 2000 and how well they perform on AI-generated codes, however we still selected one deep-learning (DL) based tool and one transformer-architecture (TA) based tool to assess the relevance of classical tools in the AI era.

In the text-based CCD category, our selection comprises Simian [54], acknowledged as a mature, popular, and classical tool for identifying straightforward code duplications [55], [56], [25]. Additionally, PMD/CPD [17] was chosen for its ability to identify literal code duplications, providing a well-rounded analysis and broader language support compared to Simian. Moving to the tree-matching approach, Deckard [28] was selected, leveraging its capability to explore structural aspects of code. As one of the most evaluated and popular tools in this category [25], we wanted to explore its performance on AI-generated clones. Within the same category, we included Clone Digger [12], an abstract syntax tree-based solution, despite its popularity not aligning with extensive evaluations in this category. Our intent is to explore its performance across diverse benchmarks. In the byte-code category, CCCD [35] stands out as a popular tool, excelling in capturing core logic and structure without being affected by variations in source code representation [65]. Shifting to the hybrid token-based and tree-based categories, we opted against NiCad [45], [15] and iClones [22], and chose CloneWorks, which is relatively new and less popular compared to NiCad and iClones, due to its superior performance compared to NiCad and iClones in previous evaluations [62], [58], [56], [55]. In the realm of graph-based tools, the relatively recent inclusion, the Stone Detector, utilizes control flow graph analysis [68], [4]. The authors of this tool found a result similar to Oreo's, which is why we aim to explore its performance on all three benchmarks. Lastly, we include ASTNN [71] and CodeBERT [19], [5]. ASTNN uses neural network architecture to learn from the abstract syntax tree and provided a strong base for AI to work on code clones. CodeBERT recognized as one of the most popular, promising, and widely used deep learning-based tools in this domain. According to the literature [19], [5], [24], its performance is similar or superior to ASTNN, FA-AST-GMN [67], CDLH [69], and RtvNN [70].

**Tool Configuration:** In the configuration of tools, we adopt a user-centric perspective, considering several factors: (1) the default settings provided by the tools, (2) the content within the documentation, and (3) the recognized characteristics of the target benchmark. These characteristics encompass clone types, syntactical similarity, and clone size. Our aim was to ensure that the recall measurements align with what an experienced user might expect when applying these tools to their own systems. Such a user would typically explore a tool's parameters and documentation, making adjustments to the default settings to suit their specific use case. Table I offers a comprehensive summary of the subject tools, the types of clones they are adept at detecting, and the configurations tailored for our benchmarks.

TABLE I: Subject Tools and Configuration

| Tools | Configuration | Tool Type | Language |
|---|---|---|---|
| PMD/CPD [17] | Minimum tokens = 50 | Text-based | Java, C, C#, Python |
| Simian [54] | Threshold = 6 | Text-based | Java, C, C# |
| Deckard [28] | Min tokens = 30, Stride = 2, Similarity = 0.7 | Tree Matching | Java |
| Clone Digger [12] | Default | Abstract Syntax Tree | Python |
| CCCD [35] | Default | Byte code | C |
| CloneWorks [58] | Type 3 - Conservative, Similarity = 0.7, Minimum lines = 4 | Token-based | Java, C, C#, Python |
| Stone Detector [4] | Default | Graph-based | Java |
| ASTNN [71] | fine-tuned on BigCloneBench | Deep Learning Model | Java |
| CodeBERT [19] | Pretrained model | Transformer Architecture based Model | Java, Python |

The configurations of the tools were primarily established by their default settings and documented recommendations. While lowering the syntactic similarity threshold or configuring the tools in a manner to boost recall, might potentially enhance their capacity to detect more clones [29], but it can lead to an increase in false positives. Given that users are inclined to use recommended thresholds, our results are a reflection of the typical usage of these tools. In instances where a setting lacked thorough documentation, we conducted experiments to observe its impact. Our approach was carefully designed to avoid over-configuring or over-optimizing the tools specifically for the benchmark, as these tools should mirror the practical constraints users face in their systems. The configurations for each benchmark were largely uniform, with the exception of variations in the minimum clone size. All the benchmarks enforced a strict minimum clone size, enabling us to configure the tools for clone size with a high degree of confidence. While alternate configurations might yield improved recall, the selected configurations represent the practices of an experienced user.

## VI. RESULTS

This section assesses tool performance using GCB (RQ1), BCB, and SCB, followed by a comparative analysis on tools' performance (RQ2). Table II presents recall, precision, and f1-score measurements from all benchmarks. We measured precision by manually validating a randomly selected set of 50 clone pairs for each tool that ran on a system. The clones of all tools are shuffled and manually tagged by judges as clones or false clones. We provided the precision results to double-check the performance of the tools in our experimental settings. Measuring recall of a tool in a subject system is more challenging because it requires knowledge of all clone pairs that naturally exist within the software system. The precision validation helps establish confidence in our overall evaluation framework.

### A. GPTCloneBench

GCB mostly consists of Moderately Type-3(MT3) and Type-4 clone pairs. We present the results of our evaluation of various CCD tools with these types of clone pairs (RQ1).

**Java Moderate Type-3 and Type-4:** From Table II, CodeBERT achieved 87% recall for Moderate Type-3 and CloneWorks demonstrated 70% recall while others exhibited lower recall rates. For Type-4 clones, the recall performance of most CCD tools was primarily lacking, with the exception of CodeBERT, which showcased 91% recall. CloneWorks fell short with 20% recall. More semantic-awareness needed to detect clones in these regions. It is more desirable for a tool to include the Type-3 regions in its detection of these clones. Otherwise, the user has to manually recognize the larger Type-3 clone. We were expecting better performance from CodeBERT and ASTNN in GCB due to its association with AI.

**C Moderately Type-3 and Type-4:** The majority of CCD tools' recalls were falling short of 50% when challenged with C Moderate Type-3 clone pairs. From Table II, CloneWorks achieved 67% recall, surpassing the performance of the other tools. For Type-4 code clones, while CloneWorks exhibited a recall rate of 28%, others were lower than that. The similar performance of CloneWorks in Java and C indicates its continuity and similar pipeline. For CCCD, usage of the Levenshtein similarity score [10] may be holding it back from attaining higher performance as CCCD only considered a score of 35. As CodeBERT did not list C as a supported language, we did not run CodeBERT on C.

**C# Moderate Type-3 and Type-4:** Similar to C, majority tools fell significantly short of achieving a 50% recall rate for Moderate Type-3 clones. Here again CloneWorks distinguished itself by achieving 73% recall. Extending the evaluation to Type-4 code clones, it is evident that the performance of all CCD tools consistently fell short of the 50%. CloneWorks once again emerged as the topper, demonstrating a recall rate of 28%.

**Python Moderate Type-3 and Type-4:** From Table II, CodeBERT and CloneWorks stand out with 87% and 71% recall respectively for Moderate Type-3. Along with CloneDigger, PMD/CPD once again came at the bottom while detecting clones from Moderate Type-3. Focusing on Type-4 code clones, the recall performance of most CCD tools remarkably lags behind, with the exception of CodeBERT, which showcased an 88% recall. Contrasting with this, CloneWorks, while achieving a 24% recall rate, falls short of matching CodeBERT's level of effectiveness.

*In summary, we expected that DL-and-TA-based tools would excel in GCB due to its association with AI. Token-based tools*

TABLE II: Recall, Precision and F1-Score Results with different benchmarks

| Tool | Language | Recall | | | | | | | Precision | F1 Score | | | | | | |
|---|---|---|---|---|---|---|---|---|---|---|---|---|---|---|---|---|
| | | GCB | SCB | BCB | | | | | | GCB | SCB | BCB | | | | |
| | | MT3 | T4 | T4 | T1 | T2 | T3 | T4 | | MT3 | T4 | T4 | T1 | T2 | T3 | T4 |
| PMD/CPD | Java | 0.19 | 0.02 | 0.029 | 1 | 0.94 | 0.31 | 0 | 0.5 | 0.28 | 0.04 | 0.055 | 0.7 | 0.65 | 0.38 | 0 |
| | C | 0.06 | 0.01 | 0.014 | | | | | 0.48 | 0.11 | 0.02 | 0.027 | | | | |
| | C# | 0.24 | 0.05 | 0.036 | | | | | 0.45 | 0.31 | 0.09 | 0.067 | | | | |
| | Python | 0.08 | 0.01 | 0.017 | | | | | 0.47 | 0.32 | 0.02 | 0.033 | | | | |
| Simian | Java | 0.16 | 0.02 | 0.027 | 0.95 | 0.78 | 0.22 | 0 | 0.52 | 0.24 | 0.04 | 0.051 | 0.67 | 0.62 | 0.31 | 0 |
| | C | 0.16 | 0.04 | 0.023 | | | | | 0.48 | 0.24 | 0.07 | 0.046 | | | | |
| | C# | 0.19 | 0.03 | 0.024 | | | | | 0.45 | 0.27 | 0.06 | 0.046 | | | | |
| Deckard | Java | 0.51 | 0.17 | 0.017 | 0.6 | 0.58 | 0.35 | 0.01 | 0.42 | 0.46 | 0.24 | 0.033 | 0.49 | 0.49 | 0.38 | 0.02 |
| Clone Digger | Python | 0.01 | 0.001 | 0.007 | | | | | 0.40 | 0.02 | 0.002 | 0.014 | | | | |
| CCCD | C | 0.31 | 0.16 | 0.059 | | | | | 0.5 | 0.038 | 0.24 | 0.106 | | | | |
| CloneWorks | Java | 0.70 | 0.20 | 0.047 | 1 | 0.99 | 0.55 | 0 | 0.97 | 0.81 | 0.33 | 0.09 | 0.98 | 0.98 | 0.7 | 0 |
| | C | 0.67 | 0.28 | 0.067 | | | | | 0.95 | 0.79 | 0.43 | 0.125 | | | | |
| | C# | 0.73 | 0.28 | 0.08 | | | | | 0.93 | 0.82 | 0.43 | 0.15 | | | | |
| | Python | 0.71 | 0.24 | 0.111 | | | | | 0.93 | 0.81 | 0.38 | 0.198 | | | | |
| Stone Detector | Java | 0.54 | 0.10 | 0.046 | 1 | 1 | 0.68 | 0.221 | 0.95 | 0.69 | 0.18 | 0.088 | 0.97 | 0.97 | 0.79 | 0.359 |
| ASTNN | Java | 0.5 | 0.31 | 0.11 | 1 | 1 | 0.92 | 0.9 | 0.54 | 0.52 | 0.39 | 0.18 | 0.7 | 0.7 | 0.68 | 0.675 |
| CodeBERT | Java | 0.87 | 0.91 | 0.766 | 1 | 1 | 0.889 | 0.945 | 0.51 | 0.64 | 0.65 | 0.612 | 0.68 | 0.68 | 0.648 | 0.662 |
| | Python | 0.87 | 0.88 | 0.839 | | | | | 0.53 | 0.66 | 0.66 | 0.65 | | | | |

like CloneWorks maintained a continuous performance. Concerning tree-based tools, we predicted they would rank third among all tool types. Most tree-based tools aligned with this expectation, except for Clone Digger. Lastly, we anticipated that text-based tools, representing an earlier generation of tools, would perform less effectively, a hypothesis that held true.

*B. SemanticCloneBench*

SCB consists of Weak Type-3/Type-4 clones for Java, C, C#, and Python. From the GCB evaluation, we have seen that it is difficult to detect Type-4 clones. In this sub-section, we will investigate how those tools performed on SCB from Table II.

**Java Type-4:** From Table II, the recall performance of the majority of CCD tools was conspicuously deficient, underscoring the complexity of this particular category and signifying the importance of more semantic awareness. Nevertheless, CodeBERT emerged as an exceptional outlier, achieving a noteworthy 76.6% recall rate, signifying a remarkable level of proficiency in the identification of such clones.

**C Type-4:** It is evident from Table II that none of the CCD tools succeeded in attaining a substantial recall rate. Particularly, CloneWorks and CCCD both yielded relatively comparable recall rates, standing at 6.7% and 5.9%, respectively. Conversely, the remaining tools exhibited even more modest recall rates.

**C# Type-4:** Similar to the findings within the C Type-4 clone category, same trend prevails within the C# Type-4 clone classification, where none of the CCD tools achieved a substantive recall rate (Table II). Here, our focus narrows to CloneWorks, which emerges as a notable outlier, having achieved the highest recall among all tools with a noteworthy 8% recall rate for C# Type-4 clones.

**Python Type-4:** A similar pattern has found within the Python Type-4 clone classification, wherein a significant recall rate was attained by none of the CCD tools, with the notable exception of CodeBERT. CodeBERT distinguishes itself by achieving an impressive 84% recall for Python Type-4 clones. Conversely, CloneWorks, while achieving a recall rate of 11.1%, could not approach the level of effectiveness exhibited by CodeBERT. In particular, the remaining tools failed to surpass CloneWorks' recall rate.

*C. BigCloneBench*

BCB consists of Java programming language. As this benchmark consists of all types of clones, we have evaluated tools with all the types of clones. We have dived into how those tools performed on BCB from Table II.

**Type-1:** PMD/CPD, CloneWorks, StoneDetector and CodeBERT have perfect Type-1 recall. Simian received an excellent recall of 95%. Deckard falls behind others with 60% recall.

**Type-2:** StoneDetector and CodeBERT have achieved a perfect Type-2 recall. CPD and CloneWorks managed more than 90% recall. Simian had a decent recall of 78% while Deckard had a poor performance with 58% recall.

**Type-3:** Within the domain of Type-3 code clones, the observation reveals that the majority of CCD tools encountered substantial challenges in achieving a recall rate surpassing 50%, with two notable exceptions, CodeBERT and CloneWorks. CodeBERT, demonstrating a particularly robust performance, excels with an 89% recall rate, and CloneWorks managed to attain a respectable 55% recall, thus presenting a noteworthy performance. Deckard and PMD/CPD, conversely, returned relatively comparable recall rates, standing at 35% and 31%, respectively. Simian exhibited a recall rate of 22%. Upon deeper examination of Type-3 clones, it becomes apparent that, within the domain of Moderate Type-3 clones, the attainment of a notable recall rate was primarily limited to CodeBERT, which distinguished itself with a high recall of 91.4%. In contrast, the remaining tools exhibited recall rates that did not reach a level worthy of report, with the exception of Deckard, which achieved a 12% recall.

**Type-4:** For Type-4 code clone pairs, it becomes evident that apart from CodeBERT and StoneDetector, the broader output of CCD tools did not produce a recall rate of noteworthy significance. Specifically, CodeBERT fine-tuned for the BCB, yielded an exceptionally high recall of 94.5%. In contrast, StoneDetector achieved a recall rate of 22% for Type-4 clone pairs.

### D. Execution Time of Each Tool

In this section, our focus was on calculating the execution time of each CCD tool for GCB. We have employed a framework, calculating the time elapsed as each tool processes a project subsequent to the injection of a single clone derived from GCB. We excluded CodeBERT and ASTNN from our analysis, because in addition to taking a huge time to get the result with our setup, it requires the user to prepare the input files that include function extraction and preprocessing to run on a project. A detailed exposition of these execution times is presented in Figure 3. Project size means the lines of code (LOC) present in that project and execution time is calculated in seconds. We have conducted this experiment on a Linux (Ubuntu 22.04 LTS) machine with 128 GB of RAM, 12th Gen Intel® Core™ i7-12700K processor and NVIDIA GeForce RTX 3080 GPU.

Turning to the Java programming language, for project sizes up to 10,000 (10k) LOC, all tools exhibit uniformity in execution times. However, as the project size expands from 10k to 100,000 (100k) LOC, with the exception of Deckard, all other CCD tools showed similar execution times. For 1 million (1M) LOC, CloneWorks, PMD/CPD, and Simian maintained their faster performance, successfully completing the task within 20 seconds.

Upon extending the investigation to C programming language, up to project sizes of 10k LOC, the performance of CCD tools bears a striking resemblance. However, as we increased the project size to 100k LOC, with the exception of CCCD which took 13x more time to process than 10k LOC, all other tools executed their tasks within 2 seconds. For 1M LOC, CloneWorks, and PMD/CPD are able to process it in about 10 seconds. However, for Simian, the processing time extends to approximately one minute under these conditions.

For C# programming language even after increasing the size to 100k LOC, there is not much difference in execution time among the tools. When it comes to a 1M LOC, even though CloneWorks and Simian took around 20 seconds, PMD/CPD managed to handle this situation within 5 seconds.

In the realm of Python, for project sizes up to 10k LOC, all tools finished processing within a matter of seconds. However, as we escalated the project size from 10k to 100k LOC, CloneWorks and PMD/CPD maintained similar performance but CloneDigger amplified its execution time by almost 52 times. For a project size of 1M LOC, PMD/CPD continues to assert its exceptional adaptability by processing the 1M-size project within 4 seconds.

## VII. DISCUSSION

### A. (RQ1) Tools Recall on GPTCloneBench

We compared and evaluated nine different CCD tools (text-based, token-based, AST-based, Byte-code-based, DL-based and TA-based) using GCB. Tools recall is shown in Table II. It is found that CodeBERT emerges as the preeminent performer among tools for Java, excelling in both Moderate Type-3 and Type-4 clones because of the association with AI. However, ASTNN didn't achieve a good recall, demonstrating its lack of context understanding compared to CodeBERT. For Java's Moderate Type-3 clone pairs, CloneWorks achieves 70% recall, signifying commendable proficiency. StoneDetector and Deckard exhibit comparable recall rates exceeding 50% within this category. This indicates that even though these tools may require more semantic awareness, they cover at least 50% of a Type-3 reference clone. It is more desirable for a tool as it helps the user to validate most of the clones. Transitioning to the C programming language, although CloneWorks attains a commendable 67% recall for Moderate Type-3 clone pairs, the rest of the tools encounter challenges in achieving notable recall rates for Type-4 clone pairs. A parallel pattern is observed in the C# programming language, wherein CloneWorks excels in Moderate Type-3 clone pairs, yet the collective performance of all tools diminishes considerably in the realm of Type-4 clone pairs. In Python, CodeBERT distinguishes itself with high recall rates for both Moderate Type-3 and Type-4 clone pairs. In comparison, CloneWorks demonstrates commendable performance exclusively for Moderate Type-3 clone pairs. For Type-4 clone pairs in Python, aside from CodeBERT, the remaining tools exhibit recall performances that do not surpass 30%.

*To sum up, from the recall measurements for the programming languages, classical CCD tools vary in their recall. However, the recently proposed tools, CloneWorks and StoneDetector, achieve a comparable recall to DL-and-TA-based tools.*

### B. (RQ2) Performance on all the benchmarks

To have a better understanding of the tools' performance on AI-generated clones and avoid any bias in evaluating DL-and-TA-based tools on AI-generated clone benchmarks, we evaluated the same tools on other benchmarks, the SCB and BCB. Results can be used to compare tools on different benchmarks, at the same time understand the differences among the benchmarks. Table II shows the recall, precision, and f1-score of tools on all three benchmarks. Results indicate that all tools perform similarly on GCB and SCB. While classical tools show good results for Type-1, Type-2, and Type-3 clones, they didn't perform well on Type-4 clones. On the other hand, DL-and-TA-based tools (CodeBERT and ASTNN) excel in detecting Type-3 and Type-4, but the precision of classical tools, namely CloneWorks and StoneDetector, is much higher than DL-and-TA-based tools. Besides, DL-and-TA-based tools need more execution time and more hardware configuration, which limit their scalability, see Figure 3.

As the three benchmarks have different types of clones, our focus is to compare benchmarks common clone types (Type-4). As Table II shows a similar trend of tools on the three code benchmarks, Tables III, and IV demonstrate recall differences among benchmarks, wherein the ∆ denotes the recall difference of tools across benchmarks. From Table III, tools show a deviation in their Moderate Type-3 recall between GCB and BCB. The deviation is clear because of the definition of the Moderate Type-3 clone in the benchmarks. While GCB classifies clones with textual similarity (50%-75%) as Moderate Type-3, BigCloneBech has a similarity range, 50%-70%, for Moderate Type-3 clones. This also explains the minimal deviation of CodeBERT as it uses AI.

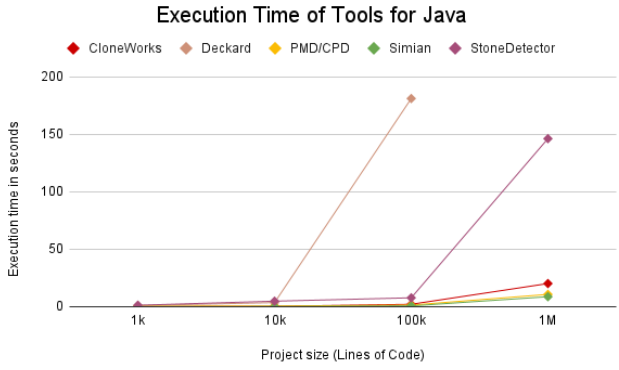
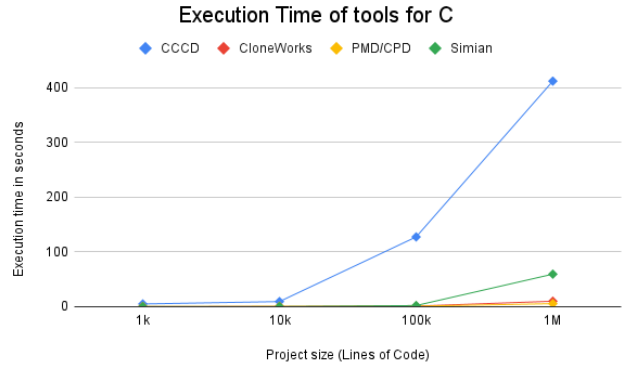
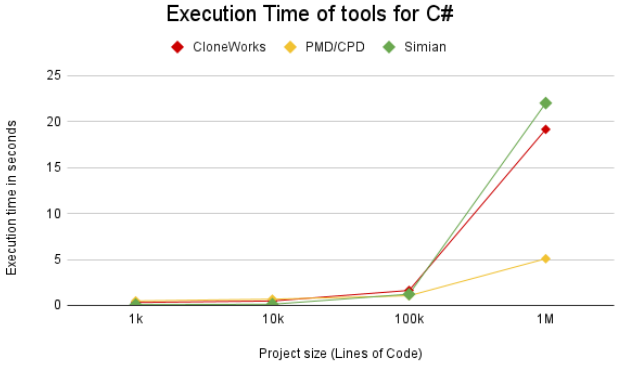
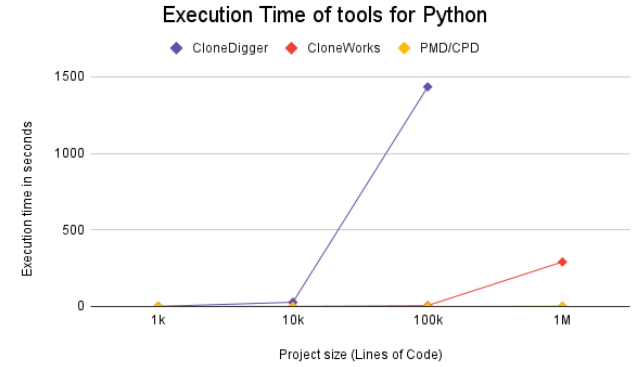

Fig. 3: Execution time of each tool for all languages using GCB

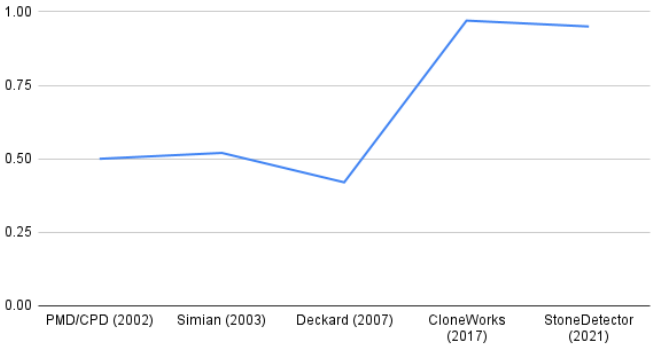

Fig. 4: Precision over time for Java

TABLE III: Recall difference of GCB, BCB, & SCB for Java

| Tool | Δ(GCB and BCB) | | Δ(GCB and SCB) | Δ(BCB and SCB) |
|---|---|---|---|---|
| | ΔMT3 | ΔT4 | ΔT4 | ΔT4 |
| PMD/CPD | 0.18 | 0.02 | -0.01 | -0.029 |
| Simian | 0.16 | 0.02 | -0.01 | -0.027 |
| Deckard | 0.39 | 0.16 | 0.15 | -0.007 |
| CloneWorks | 0.63 | 0.20 | 0.16 | -0.047 |
| StoneDetector | 0.01 | -0.12 | 0.05 | 0.175 |
| ASTNN | -0.31 | -0.59 | 0.2 | 0.79 |
| CodeBERT | -0.05 | -0.04 | 0.14 | 0.179 |

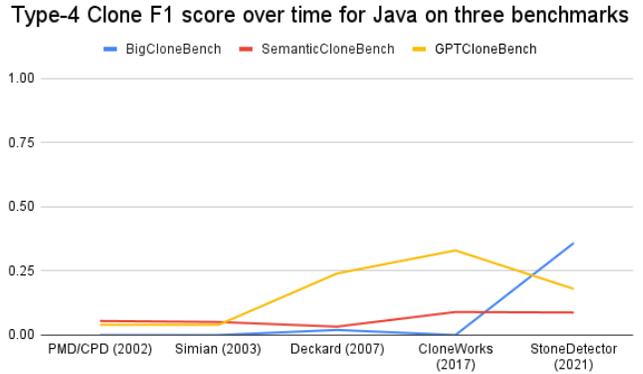

Fig. 5: F1 score over time

Transitioning to Type-4 clones, a majority of tools struggle to surpass the 50% recall threshold for all benchmarks except for CodeBERT. Especially, the recall differences in this category are lower than those observed for Moderate Type-3 clones, signifying the similarity of clones in benchmarks and highlighting the difficulty in detecting Type-4 clones, see Table III. The recall differences are not significant across GCB and SCB for all tools, with the maximum difference by ASTNN and CloneWorks. The explanation behind the minimal differences has multiple factors: Firstly, due to the fact that tools struggle with semantic clones; and lastly, semantic clones have no or very little textual or structural similarity. Similar performance is noticeable between BCB and SCB apart from ASTNN because of fine-tuning ASTNN with BCB. Furthermore, Table IV shows the recall differences between benchmarks for other programming languages. Results emphasize the fact that there is no significant recall variation for tools across semantic clone benchmarks. CCCD recorded the highest differences at 0.11 recall in C, and CloneWorks got 0.13 recall for Python, which is still not significant [56].

TABLE IV: Recall difference of GCB & SCB for C, C# and Python

| Language | Tool | ΔGCB and SCB ΔT4 |
|---|---|---|
| C | PMD/CPD | -0.01 |
| C | Simian | 0.02 |
| C | CCCD | 0.11 |
| C | CloneWorks | 0.22 |
| C# | PMD/CPD | 0.02 |
| C# | Simian | 0.01 |
| C# | CloneWorks | 0.2 |
| Python | PMD/CPD | -0.01 |
| Python | CloneDigger | -0.01 |
| Python | CloneWorks | 0.13 |
| Python | CodeBERT | 0.04 |

Our evaluation of classical CCD tools revealed that CloneWorks and StoneDetector outperformed other classical approaches, with CloneWorks demonstrating the highest effectiveness. In our study, we experimented with both conservative and aggressive normalization strategies of CloneWorks. Both approaches represent code fragments as structured sets of code-statement patterns while normalizing identifiers and literal values within those statements. However, the conservative approach, which tokenizes terms rather than relying on line-based segmentation, consistently yielded better performance. This finding underscores the critical role of effective normalization techniques in enhancing classical CCD tool performance, particularly when detecting AI-generated clones, where variability in syntax and structure is more noticeable. Our results emphasize that well-designed normalization strategies can significantly improve clone detection accuracy.

As classical CCD tools strive to achieve better performance, we displayed the tools' performance evolution over time for Java, as most of the tools and benchmarks have support in Java. Figure 4 illustrates the evolving precision of classical clone detectors over time. It highlights that the StoneDetector, the most recent tool, achieves an impressive precision of nearly 95%. This is noteworthy, especially when compared to tools like CodeBERT and ASTNN. However, post-2016, there was a noticeable shift towards the development of more ML or DL or TA-based tools [38], [36], [51]. This shift is evident when comparing the impact of tools like CodeBERT, launched in 2020, to classical tools such as CloneWorks, released in 2017, where we can find that CodeBERT is comparatively more popular in the code-clone community. Furthermore, Figure 5 presents the overall F1 scores for classical tools, specifically targeting Type-4 clones within the Java programming language, as Type-4 clones are the most difficult to detect, and AI-generated codes are mostly semantic clones. We can see that over time, the F1-score for Type-4 clones has slightly improved for classical tools. Nevertheless, our evaluation suggests that classical tools still hold significant promise for detecting semantic or AI clones.

*Results of RQ2 indicate that despite DL-and-TA-based tools achieving a better recall in Type-3 and Type-4 clones, classical tools can still perform better in terms of precision, execution time, and scalability. Also, among all the classical CCD tools, CloneWorks with conservative approach provided the better performance, highlighting the effectiveness of having a well-designed normalization techniques in enhancing clone detection accuracy. From the user's perspective, considering the results of AI detection tools means tolerating the risk of having false positives that might affect the results' generalizability. When it comes to clone detector selection, one must keep in mind that classical CCD showed better performance considering all the evaluation metrics for detecting Type-1, Type-2, and Type-3 clones. However, when it comes to Type-4 clones, even though the recall is not as high as that of DL-and-TA-based tools, the precision is much higher. Users must accept this trade-off while selecting a tool for CCD. Common wisdom posited that there would be minimal differences in the recall results between GCB and SCB because GCB was constructed utilizing SCB as a foundational element. The results of our evaluations align with this hypothesis, revealing a notable consistency between the two benchmarks. Also, the recall of tools with GCB is within the range of BCB Type-4 Recall.*

### C. (RQ3) Execution time and scalability

To summarize the findings from Section VI-D, *CloneWorks and PMD/CPD often joined by Simian, demonstrated consistently fast execution times across all programming languages, including on projects up to one million LOC, demonstrating their scalability and efficiency, which helps to explain their continued popularity in the industry.* So, when scalability is a critical constraint, these tools remain the most practical and operationally viable choice. In contrast, other tools struggled to maintain acceptable performance levels as project size increased.

## VIII. THREATS TO VALIDITY

The performance of CCD tools can vary considerably based on the specific configurations employed. Wang et al. [66] described this phenomenon the "confounding configuration choice problem," a challenge that extends through the area of clone studies. In our research, we took deliberate measures to address this challenge by carefully selecting tool configurations based on tool documentation and benchmark alignment, not dataset-specific tuning that align with the known properties of the benchmark, including clone types and clone size. Where benchmark-specific insights were lacking, we leaned on the default settings and recommendations of the tools while integrating our extensive knowledge of the benchmarks into the decision-making process. This approach closely mirrors the steps a typical user would take to configure a tool for their own system, ensuring that our results provide a reflection of user expectations. Importantly, we refrained from exhaustive, trial-and-error testing of the tools across various settings, recognizing the impracticality of such an approach for users. It's noteworthy that in the case of Type-3 clone detectors, reducing their similarity thresholds might increase the number of clones detected [29]. Addressing potential biases in code fragment selection for precision measurement is crucial in CCD research. To mitigate this issue, our methodology incorporates a random selection of code fragments, ensuring a representative sample across diverse subject systems. Additionally, experienced software developers with several years of expertise were engaged as judges to cross-verify the results, further enhancing the validity and reliability of our precision assessment. This approach aims to minimize selection bias and improve the robustness of our findings. Furthermore, another threat that can arise after the injection of the clone pair is whether the evolved system architecture changes and whether it is still compilable and executable. We solved the issue by making sure when we inject the pair, they are injected into the proper place of the subject system with existing functions, and they won't hamper the overall architecture, coding style, indentation, and other related issues. To reduce sampling bias, we evaluated each detector on the full set of within-language clone pairs from GCB, SCB, and BCB, without applying any additional filtering or selection. This ensures that our results reflect the original distributions of clone types, programming languages, and project contexts defined by the benchmark creators. While this approach avoids bias from selective sampling, it also means that performance may vary on datasets with very different characteristics. To mitigate this threat, we selected three distinct and widely used benchmarks, each covering different aspects of clone detection. This diversity helps balance out benchmark-specific characteristics, and while some variation may still exist, we believe the impact is limited given the consistency observed across all three benchmarks.

## IX. Conclusion

AI-generated code has become an integral part of modern software development resulting in an increase in AI-generated clones. Despite the efficiency, scalability, and robustness, classical CCD tools have not been systematically evaluated on AI-generated clones, raising concerns about their effectiveness in this AI era. Our evaluation reveals that classical CCD tools retain the potential to detect AI-generated clones, particularly when leveraging effective normalization techniques. Notably, CloneWorks, with its conservative normalization approach, demonstrated superior performance, underscoring the critical role of structured normalization. However, detecting Type-4 clones remains a challenge, emphasizing the need for further refinement of classical techniques to handle the unique systematic variability present in AI-generated code. Moreover, our finding reinforces that BCB alone is insufficient for evaluating modern clone detectors, as it lacks AI-generated clones and some ambiguity regarding Type-4 clones. GCB and SCB provide critical data for assessing tools across human-written and AI-generated clones in multiple languages. In conclusion, while TA-and-DL-based clone detectors are advancing, classical tools remain highly relevant due to their efficiency, and wide adoption. To ensure their continued effectiveness in the AI era, it is essential to evaluate, refine, and enhance classical detectors to address AI-generated clones, maintaining their role in software quality.

## Acknowledgment

This research is supported in part by the Natural Sciences and Engineering Research Council of Canada (NSERC) Discovery Grants program, the Canada Foundation for Innovation's John R. Evans Leaders Fund (CFI-JELF), and by the industry-stream NSERC CREATE in Software Analytics Research (SOAR).